# High-resolution ptychographic imaging at a seeded free-electron laser source using OAM beams

**Matteo Pancaldi,**[1,2,†,*] **Francesco Guzzi,**[1,15,†] **Charles S. Bevis,**[3] **Michele Manfredda,**[1] **Jonathan Barolak,**[4] **Stefano Bonetti,**[2,5] **Iuliia Bykova,**[6] **Dario De Angelis,**[1] **Giovanni De Ninno,**[1,7] **Mauro Fanciulli,**[8,9] **Luka Novinec,**[1] **Emanuele Pedersoli,**[1] **Arun Ravindran,**[7] **Benedikt Rösner,**[6] **Christian David,**[6] **Thierry Ruchon,**[8] **Alberto Simoncig,**[1] **Marco Zangrando,**[1,10] **Daniel E. Adams,**[4] **Paolo Vavassori,**[11,12] **Maurizio Sacchi,**[13,14] **George Kourousias,**[1] **Giulia F. Mancini,**[3] **and Flavio Capotondi**[1]

[1]*Elettra-Sincrotrone Trieste S.C.p.A., 34149 Basovizza, Trieste, Italy*
[2]*Department of Molecular Sciences and Nanosystems, Ca' Foscari University of Venice, 30172 Venezia, Italy*
[3]*Laboratory for Ultrafast X-ray and Electron Microscopy (LUXEM), Department of Physics, University of Pavia, 27100 Pavia, Italy*
[4]*Department of Physics, Colorado School of Mines, 1523 Illinois Street, Golden, Colorado 80401, USA*
[5]*Department of Physics, Stockholm University, 10691 Stockholm, Sweden*
[6]*Paul Scherrer Institut, 5232 Villigen-PSI, Switzerland*
[7]*Laboratory of Quantum Optics, University of Nova Gorica, 5001 Nova Gorica, Slovenia*
[8]*Université Paris-Saclay, CEA, CNRS, LIDYL, 91191 Gif-sur-Yvette, France*
[9]*Laboratoire de Physique des Matériaux et Surfaces, CY Cergy Paris Université, 95031 Cergy-Pontoise, France*
[10]*Istituto Officina dei Materiali, CNR, 34149 Basovizza, Trieste, Italy*
[11]*CIC nanoGUNE BRTA, Tolosa Hiribidea, 76, 20018 Donostia-San Sebastián, Spain*
[12]*IKERBASQUE, Basque Foundation for Science, Plaza Euskadi 5, 48009 Bilbao, Spain*
[13]*Sorbonne Université, CNRS, Institut des NanoSciences de Paris, INSP, 75005 Paris, France*
[14]*Synchrotron SOLEIL, L'Orme des Merisiers, Saint-Aubin, B. P. 48, 91192 Gif-sur-Yvette, France*
[15]*francesco.guzzi@elettra.eu*
[†]*These authors contributed equally to this work*
[*]*matteo.pancaldi@elettra.eu*

**Abstract:** Electromagnetic waves possessing orbital angular momentum (OAM) are powerful tools for applications in optical communications, new quantum technologies and optical tweezers. Recently, they have attracted growing interest since they can be harnessed to detect peculiar helical dichroic effects in chiral molecular media and in magnetic nanostructures. In this work, we perform single-shot per position ptychography on a nanostructured object at a seeded free-electron laser, using extreme ultraviolet OAM beams of different topological charge order $\ell$ generated with spiral zone plates. By controlling $\ell$, we demonstrate how the structural features of OAM beam profile determine an improvement of about 30% in image resolution with respect to conventional Gaussian beam illumination. This result extends the capabilities of coherent diffraction imaging techniques, and paves the way for achieving time-resolved high-resolution (below 100 nm) microscopy on large area samples.

## 1. Introduction

Maxwell's equations show that electromagnetic radiation carries energy, polarization and angular momentum. While the first two properties have extensively been used in several experimental approaches, only recently the latter quantity, i.e. the specific spatial distribution of the electromagnetic field phase related to the orbital angular momentum (OAM) of light, has emerged as a valuable experimental tool. The first observation of OAM carried by the electromagnetic field was reported by Allen et al. [1] in 1992. In their work, the authors showed that light described by an $\exp(i\ell\phi)$ spiraling phase structure intrinsically possesses an OAM equivalent to $\ell\hbar$ per photon, where the parameter $\ell$ defines the so-called "topological charge order" of the beam ($\ell \in \mathbb{Z}$). After this groundbreaking work, OAM light quickly unlocked new areas of research due to its peculiar applications in both fundamental and applied physics [2–4]. Nowadays, OAM beams are harnessed in a great variety of applications covering different fields, such as the study of dichroic and chiroptical materials [5–12], manipulation and trapping of particles [13,14], optical communication systems [15], quantum technologies [16,17], and imaging [18].

Regarding microscopy applications, OAM beams have been proposed as a tool for enhancing the edges detection in phase contrast microscopy [19–21], and for increasing the spatial resolution [22–25]. It has been proven that, at optical wavelengths, OAM beams enhance the image quality in phase-sensitive techniques, providing a uniform contrast at the interface between different elements, dislocations or morphological height variations [18,26,27]. Basically, in such experiments a spiral phase plate in a 4-f optical system acts as a bidimensional Hilbert filter in Fourier space, generating a contrast enhancement at the edges of the target structure. This is of particular importance in low-transmission contrast samples like biological cells, where topological height information can be retrieved through the analysis of the phase contrast [18,28]. At shorter wavelengths, diffraction imaging techniques based on beam coherence are routinely employed to obtain high-resolution diffraction-limited images of nanostructured samples [29–31]. Among them, ptychography represents a complete *in-situ* characterization tool, since it allows retrieving both the transmission and phase contrast image of the sample, as well as wavefront-sensing information of the illumination function [32–35]. In combination with OAM beams, ptychography has been exploited in the characterization of the vortex beam structure, and in the determination of the topological charge order [36–39]. More recently, beneficial effects given by the structured illumination of OAM beams for the ptychographic imaging of extended samples have been demonstrated by Eschen et al. [40] and Wang et al. [41]. In such cases, the enriched spatial frequency content of the OAM illumination function optimizes the distribution of scattered photons at the detector plane, relaxing the dynamic range requirements and improving the signal-to-noise ratio (SNR). Moreover, due to the reduced scattering symmetry and the increased diversity of collected diffraction patterns, the convergence and the robustness of ptychographic reconstruction algorithms are improved with respect to the case of conventional Gaussian illumination.

In this work, we report on high quality ptychography image reconstructions achieved by exploiting zone-plate-generated OAM beams [42,43] at an extreme ultraviolet (EUV) seeded free-electron laser (FEL) source [44]. With respect to previous studies [45–47], we experimentally observe that the improved pointing stability introduced by the zone plate optics, combined with the high intensity and wavelength stability of a seeded FEL source, allow us to obtain high quality images of a nanostructured Siemens star test object with single-shot per position illumination. Moreover, an increase in the achievable spatial resolution of the reconstructed ptychography images is observed at high topological charge orders, due to the structured illumination of OAM beams. This result extends the ability of coherent diffraction imaging techniques to perform time-resolved high-resolution imaging of large area samples. Finally, we observe that the retrieved illumination function can be used for determining the main residual aberrations induced by the focusing optical system. Besides the importance for imaging experiments, our analysis is also relevant for dichroic experiments involving OAM beams [10,12], since modifications in the expected intensity pattern for opposite $\ell$ values can introduce artifacts in the evaluation of the dichroic signal.

## 2. Ptychography and experimental setup

### *2.1 The ptychographic technique*

Ptychography merges the main advantages of lensless imaging [48], since it combines the use of phase retrieval algorithms with a scanning probe approach typical of x-ray microscopy [49], in order to reconstruct diffraction-limited morphological information of extended samples. An essential requirement of ptychography is the spatial overlap (typically $> 70\%$) among the areas of the sample generating single diffraction patterns. This redundancy in collected data information allows for lifting the isolated object constraint of traditional coherent diffraction imaging [34]. Being a computational imaging technique, ptychography requires an iterative reconstruction procedure, which has to be carefully checked and tuned in order to avoid artifacts. The net result is a high-resolution complex-valued image (transmission amplitude and phase) of the sample and a refinement of the probing beam illumination function [35]. As a consequence, ptychography finds applications not only as an imaging technique, but also as a wavefront-sensing metrological diagnostic tool [50–52]. Figure 1(a) shows a sketch of a typical experimental setup: the scattered beam is measured in transmission geometry, recording the diffraction patterns at the detector plane jointly with the sample coordinates. After scanning the desired set of sample positions, an iterative procedure solves the inverse problem of finding the object and the average illumination that better approximate the set of measured scattering patterns. The accuracy and convergence of a ptychography reconstruction rely on the precise knowledge of the input parameters which describe the wave propagation in the reconstruction algorithm (i.e., pulse's central wavelength and bandwidth, beam size, pixel size, propagation distances and sample positions).

Our ptychography algorithm, described in Ref. [53], automatically factorizes the underlying object and the common illumination shed on it [32,33], which can also include optical aberrations, like astigmatism or coma. Moreover, the trade-off between photon flux and spatial coherence of the illumination can be relaxed, considering partially coherent beam conditions, e.g. when multiple propagating modes incoherently add at the detector plane, which would result in a blurred diffraction pattern [54]. In this case, to solve the inverse problem, probe decomposition techniques (such as the one presented in Ref. [55]) are employed to computationally disentangle the weighted contribution of each individual mode to the recorded diffraction pattern dataset. For single-shot per position experiments, the decomposition of the illumination function in multiple modes intrinsically allows for estimating a lower bound for the degree of transversal coherence of the beam [56,57].

In order to improve the object self-consistency during the iterative process [58], we impose a refinement of the sample positions associated with each diffraction pattern in the scan [53,59]. This additional optimization procedure allows for the evaluation of spurious vibrations on the beam pointing in terms of position offsets, providing a quantitative measure of the setup stability, and it is of paramount importance for the final image quality. Similar iterative sample-position refinement approaches have also been employed in Refs. [45,46] for enforcing a common illumination, even in the extreme case of strong pointing instability due to a vibrating beam.

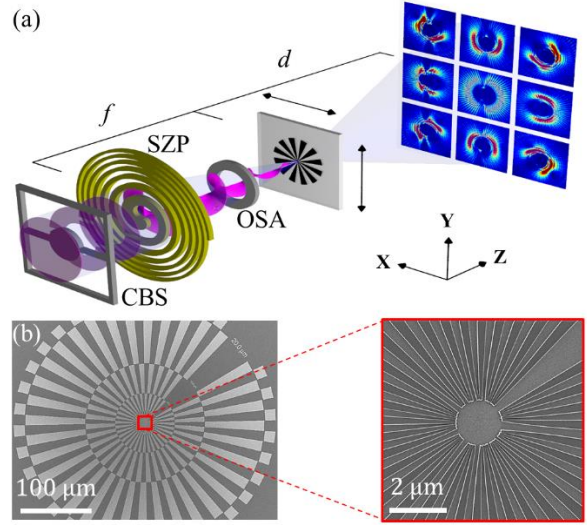

Fig. 1. (a) Schematics of the experimental setup for ptychography. The OAM beam is generated and focused on the sample plane (at a distance $f$) using a spiral zone plate (SZP). The order sorting aperture (OSA) allows only light from the first diffraction order to reach the sample, while the higher orders are blocked. Upstream to the SZP optics, a central beam stop (CBS) blocks the direct beam that does not carry OAM. The sample is then scanned in the XY-plane in order to record a set of diffraction patterns with a CCD camera (placed at a distance $d$), to be used for the ptychography reconstruction of both the object and the illumination functions. (b) Scanning electron microscopy image of the Siemens star used as a test transmitting sample, and magnification of the central part, where the smallest features have a size of 100 nm.

## *2.2 Description of the experimental setup*

The imaging capabilities of ptychography with OAM beams have been experimentally investigated at the DiProI beamline [60] of the FERMI FEL source [44]. The end-station is equipped with an active focusing system based on Kirkpatrick-Baez (K-B) mirrors [61], which were set in flat-mirrors configuration for compensating the natural divergence of the FEL source and producing a nearly collimated beam. The focusing of the EUV radiation and the generation of OAM beams were performed by means of nanofabricated spiral zone plates (SZPs) [42,43] directly placed inside the main vacuum chamber of the end-station. The FEL light was linearly polarized in the horizontal plane with a wavelength $\lambda$ of 18.9 nm (65.6 eV). Taking into consideration the beamline transmission and the 500 nm-thick Zr solid-state filter used for attenuating the total source intensity, we estimate a pre-focusing fluence of about 0.1 µJ/cm$^2$ irradiating the SZPs, well below their damage threshold and guaranteeing long-term stability.

A sketch of the experimental setup is shown in Fig. 1(a). Motorized piezo stages were used for precisely switching between seven SZPs for the generation of $\ell = 0, \pm1, \pm2, \pm3$ OAM beams [62]. All SZPs were fabricated with nominally identical physical parameters [63], in order to provide a focal length $f$ of 164.7 mm for $\lambda = 18.9$ nm radiation, with an estimated first order focusing efficiency of about 12% [43]. According to the average input fluence and to the theoretical full width at half maximum spot size (1.67 µm), we calculate a fluence of about 15 mJ/cm$^2$ at the focal plane for the Gaussian illumination case ($\ell = 0$). An order sorting aperture and a central beam stop were used for spatially filtering the higher diffraction orders of the SZPs and for blocking the unfocused direct beam, respectively.

Ptychography scans were performed on a 100 nm-thick hydrogen silsesquioxane (HSQ) Siemens star structure [64] patterned on top of a 200 nm-thick silicon membrane. The outer diameter of the Siemens star is approx. 500 µm, while the smallest features close to the center have a size of 100 nm, as shown in Fig. 1(b). The sample was mounted on a five-axis piezo goniometer, which allows for controlled movements with 100 nm of resolution. For each $\ell$ value, the center of the Siemens star was scanned in a 7×7 grid, where the grid points were chosen according to a Fermat spiral pattern with a step size of 1 μm [65]. At each point of the scan, a single FEL shot was used for illuminating the sample, and the diffracted intensity was recorded by an in-vacuum Princeton MTE2048 CCD camera (2048 × 2048 pixels, with a 13.5 μm pixel size), placed at a distance $d$ of 140 mm from the sample plane and operated in 4×4 binning mode.

## 3. Results

### 3.1 Ptychography reconstructions at a seeded FEL

The ptychographic reconstructions were carried out using the M-rPIE algorithm [66] implemented in the SciComPty software package [53], by including the simultaneous propagation of three independent illumination modes [55]. For each $\ell$ value, three ptychographic scans were recorded, and representative reconstructions are shown in Fig. 2. Each panel is composed by amplitude and phase of the main illumination mode, and by the corresponding object amplitude. The reconstructed pixel size can be precisely calibrated in order to retrieve the beam and object sizes at the sample plane, as reported in Supplement 1, Sec. S1.

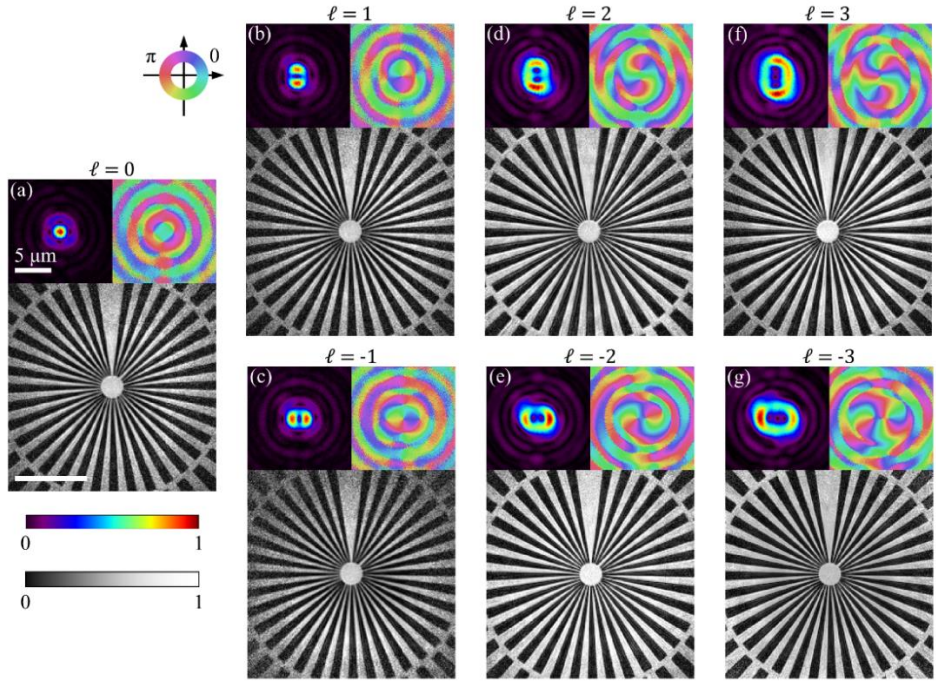

Fig. 2. (a)-(g) Ptychographic reconstructions of the Siemens star for $\ell = 0, \pm 1, \pm 2, \pm 3$. For each panel, the main illumination amplitude, the main illumination phase and the object amplitude are shown. Both the white markers in (a) correspond to a length of 5 μm, and they differ by a factor of two.

Since beam coherence is a fundamental property of FEL sources, Fig. 3(a) shows the modal weights associated to the three illumination modes for each reconstruction. The modal weight $\beta_i$ is obtained by normalizing the integrated intensity of the $i$-th mode with respect to the total intensity. Most of the weight (82% in average) is associated with the first decomposition mode, while each one of the higher order modes contributes to less than 12% to the total illumination function, hence allowing us to consider only the reconstruction of the main illumination modes in Fig. 2. As discussed in Ref. [57], the decomposition of the illumination function in orthogonal modes provides insight into the factors affecting the speckle visibility in the diffraction patterns, which can be related to, e.g., partial transverse and longitudinal coherence of the radiation, vibration of the beam around an average position, detector point spread function. It is worth pointing out that, in single-shot per position experiments, the speckle visibility degradation is a representative measurement of the beam partial coherence since, in contrast to multiple-shot exposure, the blurring of a single-shot diffraction pattern cannot be affected by beam pointing fluctuations. As a consequence, it is possible to evaluate the degree of transversal coherence $\zeta$ for the beam through the modal weights, according to $\zeta = \sum_i \beta_i^2 / (\sum_i \beta_i)^2$ [67]. Since the modal decomposition can favor the formation of unphysical modes that spread part of the diffraction-uncorrelated intensity outside the object area [57] (see Supplement 1, Sec. S2), we can only provide a lower boundary for $\zeta$. Considering the average $< \beta_i >$ values in Fig. 3(a), we obtain that $\zeta = (68 \pm 2)\%$. Such lower boundary for the degree of transversal coherence is in good agreement with previous estimates obtained on a seeded FEL source by means of the Young's double slit interference experiment [44,60]. In a recent work, Kharitonov et al. [46] performed a similar analysis using a self-amplified spontaneous emission (SASE) FEL source [68]. They reported, in a comparable spectral range, a transversal coherence of about 54%, and the need for nine independent illumination modes to obtain reliable and stable reconstructions of the object. Our mode decomposition analysis shows favorable conditions in terms of transversal coherence quality for seeded FEL radiation in comparison to a SASE FEL source. Moreover, the seeding scheme also provides a higher degree of longitudinal coherence with respect to SASE, with the generated radiation behaving statistically as a laser [69]. As a consequence, the narrow spectral bandwidth of seeded FEL radiation allows for a well-defined focal length for the chromatic SZP optical element, so increasing the number of photons available inside the coherent scattering volume and improving the contrast visibility of the speckle pattern. These properties make the seeded FEL an ideal source for performing time-resolved high-resolution ptychography experiments.

Besides beam coherence, ptychography reconstructions are very sensitive to positioning errors. Beam positioning errors can be a severe challenge for stable and reliable reconstruction algorithms, especially for ptychographic imaging experiments at FEL sources due to intrinsically larger pointing instabilities with respect to synchrotron radiation facilities [45,46]. Moreover, in single-shot per position experiments the preservation of a precise positioning during the sample scan is crucial, since fluctuations are not mitigated by averaging over many shots at the expense of a worse image resolution. An estimate of our system stability can be obtained by analyzing the output of the position-refinement step of our algorithm [53]. In this procedure, the position of each diffraction pattern of the scan is optimized from its starting point to iteratively extract the offset along the scanning axis for improving the convergence of the ptychographic reconstruction. Figure 3(b) shows the distribution of the final refinement offsets in the X-Y sample plane for all sample positions considered during the experiment (approx. 1000 points). The full widths at half maximum of both the X- and Y-projection are about 300 nm. This analysis clearly shows that, due to the large numerical aperture, the use of diffractive optical elements to focus a collimated beam strongly mitigate the detrimental effects of pointing instabilities at the imaging plane by more than one order of magnitude with respect to previously performed experiments at FEL sources, where conventional reflective K-B optics were used [45,46].

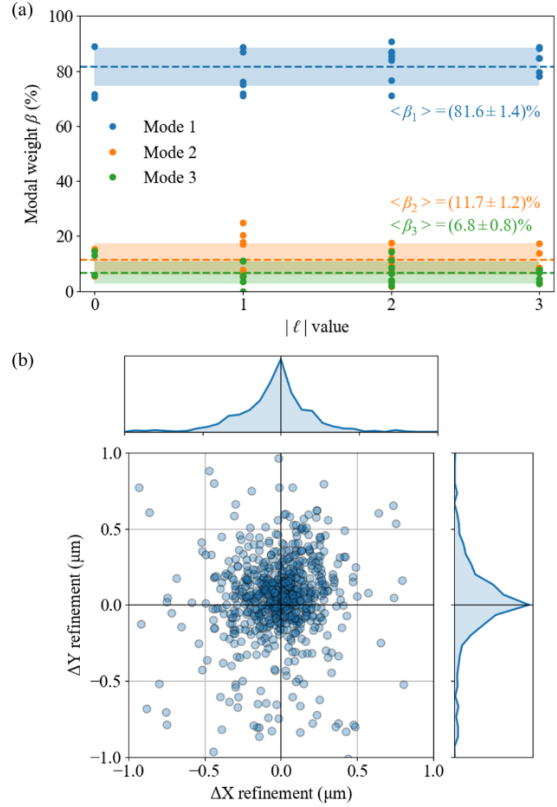

Fig. 3. (a) Modal weights $\beta_i$ associated to the three illumination modes for each reconstruction. The dashed lines highlight the average $\beta_i$ values, while the shaded areas represent the standard deviations. (b) Offsets in the scan positions as determined by the position-refinement algorithm along the X- and Y-axis for all the reconstructions. The full widths at half maximum of both the projections in the lateral plots are 300 nm.

*3.2 Analysis of the reconstructed illumination function*

In the previous section we have shown that most of the intensity of the illumination function is concentrated in the first mode of the modal decomposition. For each $\ell$ value, Fig. 2 reports the corresponding principal mode illumination, showing that the ptychographic reconstructions successfully identify the correct phase winding direction and topological charge order of the selected SZP. Indeed, by looking at the central part of the figure panels reporting the illumination profiles, a $2\pi$ phase winds $|\ell|$ times around the propagation axis, and the direction is opposite for opposite signs of the topological charge order. However, a careful inspection of all the retrieved spots shows two principal distortions with respect to the ideal "donut-like" shape of a Laguerre-Gaussian mode [1,42]: (i) an azimuthal modulation of the beam intensity and (ii) a separation of the expected vortex singularity in multiple poles for topological charges $|\ell| > 1$. With the help of simulations (see Supplement 1, Sec. S3), we were able to identify the 45-degree astigmatism induced by the beamline K-B optics as the major source of the above-mentioned probe beam deformations. The effect of this optical aberration is to split the central phase singularity of an ideal Laguerre-Gaussian beam into $|\ell|$ single-order singularities aligned along a line [39,70,71]. Despite the presence of astigmatism, the total OAM topological charge order is conserved, as shown by the proper winding of the overall reconstructed phase and as discussed in Ref. [72].

The presence of 45-degree astigmatism was confirmed by numerically propagating the reconstructed $\ell = 3$ illumination function along the Z-axis by means of the free-space propagator for a monochromatic wave [73]. Figure 4(a),(b) shows sections of this free-space propagation corresponding to the XZ-plane and to the YZ-plane, respectively. In these projected views, the beam waist is at the same position, approximately at $Z = -200$ μm, showing the perfect matching between tangential and sagittal focusing planes. However, along planes rotated by 45° with respect to the X-Y coordinate system, a relative difference of 400 μm in the beam waist position is observed (see Fig. 4(c),(d)), confirming the presence of a residual 45-degree astigmatism not fully compensated by the K-B optics.

It is worth noting that, since distortions induced by astigmatism have a different impact on the illumination function for opposite $\ell$ values, experiments based on OAM chiral dichroism [10,12] must handle with great care the quantitative comparison between measurements obtained for opposite $\ell$ values. Indeed, spurious chiral signals can be generated (e.g.) by spatial intensity variations due to optical aberrations. Ptychography, with its wavefront-sensing capabilities, could represent the ideal metrological technique for *in-situ* beam characterization, in order to minimize optical aberrations before performing the target chiral experiments.

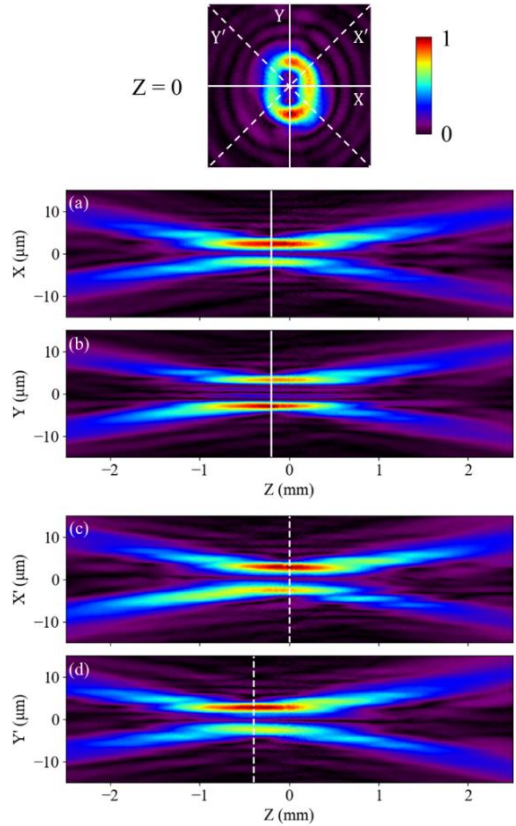

Fig. 4. Free-space propagation of the $\ell = 3$ main illumination mode. (a) Section of the free-space propagation corresponding to the XZ-plane (top view). The white line highlights the position of the beam waist. (b) Section of the free-space propagation corresponding to the YZ-plane (side view). The white line highlights the position of the beam waist, which is the same as for (a). (c) Section of the free-space propagation corresponding to the X'Z-plane. (d) Section of the free-space propagation corresponding to the Y'Z-plane. The sections in panels (c) and (d) correspond to planes rotated by 45° with respect to the X-Y coordinate system. The fact that in these views the beam waist is in different positions, as marked by the white dashed lines, confirms the presence of 45-degree astigmatism.

*3.3 Analysis of the reconstructed object image*

The intrinsic structuring of an OAM beam and the additional contribution to the illumination diversity provided by astigmatism are instrumental in enhancing the quality of the reconstructed object images. Figure 2 already shows that the fine details in the central part of the Siemens star can readily be identified, allowing to estimate a resolution comparable with the pixel size (96 nm), an order of magnitude better with respect to a SASE FEL source [46,47] and comparable with more stable sources, like table-top lasers for high-harmonic generation [40,41,74] and synchrotron radiation facilities [75,76]. To obtain a quantitative evaluation of image resolution, Fig. 5(a) shows the experimental average Fourier ring correlation (FRC) curves calculated averaging the FRC results obtained by permuting pairs of different reconstructed images collected at a given topological charge order $|\ell|$. As described in Ref. [77], the image resolution can be extracted by defining a threshold for the FRC values, which allows getting the maximum spatial frequency to achieve a predefined SNR. By considering a SNR of 0.41 (corresponding to the common 0.5-bit threshold), we obtained a resolution of $(118 \pm 7)$ nm, $(128 \pm 8)$ nm, $(96 \pm 5)$ nm, and $(85 \pm 5)$ nm for $|\ell| = 0, 1, 2, 3$, respectively. Under similar experimental conditions, a clear improvement of about 30% in the image resolution is observed for the SZP with the highest topological charge order ($|\ell| = 3$) with respect to conventional Gaussian beam illumination ($\ell = 0$). Similar beneficial effects in imaging resolution going from Gaussian illumination to OAM beams have recently been reported by Tan et al. [78] at optical wavelengths and by Eschen et al. [40] in the EUV regime. We ascribe this resolution improvement to the beam structuring and to its consequent enhancement of the probe spatial-frequency spectrum. On one hand, this effect increases the SNR due to a more efficient distribution of scattered photons at the detector plane; on the other, it increases the diversity of the recorded diffraction patterns during the scan, speeding up the reconstruction convergence.

Besides the resolution improvement due to the higher frequency content of structured beams [75,78], the effect of an OAM beam illumination is also visible at low spatial frequencies (below 0.1 in our case, see Fig. 5(a)), where the FRC values grow along with the value of $|\ell|$. According to Odstrčil et al. [79], the observed enhancement is due to the improvement in the reconstruction accuracy of the object phase, which is determined by the minor sensitivity to unwrapped phase errors with respect to constant phase probes when the size of the investigated object is larger than the illumination spot size (i.e. at low reconstructed frequencies). In other words, the OAM phase gradient at length scales comparable to the beam spot size allows to unequivocally determine the object phase among successive scanning positions during the reconstruction process, so improving the image quality at low spatial frequencies. This enhancement of image quality at spatial frequencies comparable with the beam spot size agrees with the results reported by Li and Li [23]. Indeed, their work suggests that OAM light can exceed the Rayleigh limit and achieve super-resolution when the beam is tightly focused to a spot size comparable to the radiation's wavelength.

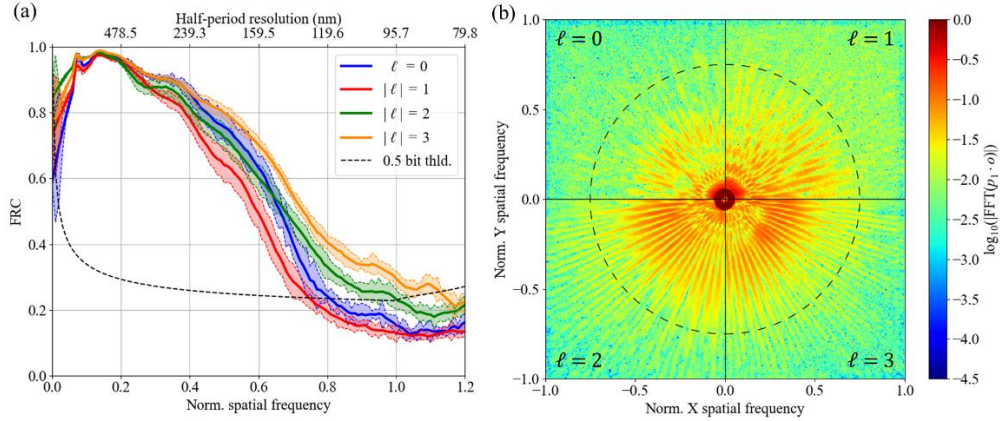

Fig. 5. (a) Average Fourier ring correlation (FRC) curves calculated on the experimental ptychography reconstructions. The upper horizontal axis shows the corresponding half-period resolution calibrated with the reconstructed pixel size. The shaded areas represent the standard deviation calculated over all the datasets illuminated with the same $|\ell|$. The black dashed line indicates the 0.5-bit threshold, which is a common criterion for defining the image resolution and corresponds to a SNR of 0.41. (b) Amplitude of the Fourier transform of the product between the main mode ($p_1$) of the reconstructed illumination functions in Fig. 2 and the average reconstructed object $o$. For comparison, each quadrant is associated to a different $\ell$ value and it is normalized to its average amplitude in the central part of the pattern. The black dashed circle has a radius of 0.75.

Despite the highest resolution being associated with the highest $\ell$ value, the $|\ell| = 1$ resolution falls out of the general trend, suggesting the existence of a trade-off between the increase of the high frequency spectral content induced by non-ideality of the illumination function and the minimum resolvable separation between two adjacent structures [78]. We ascribe this evidence to the fact that, even if the resolution should increase with $|\ell|$ due to the higher frequency content of the illumination function, the variations in the illumination function profile should match the peculiar object features to be efficient in terms of SNR. In other words, when the characteristic spatial frequency of the probing beam does not match with the minimum sample feature, the reduced speckles contrast of the far field diffraction pattern produces a deterioration of image quality. To corroborate this point, Fig. 5(b) shows the amplitude of the Fourier transform of the product between the reconstructed illumination functions in Fig. 2 (for $\ell = 0, 1, 2, 3$) and the average reconstructed object. From this quantity, which in logarithmic scale is proportional to Fraunhofer diffraction [73], it is indeed possible to compare the contrast of the far field speckle pattern for different $\ell$ values. It is apparent that, at high spatial frequencies (e.g. above 0.75, see the black dashed circle in Fig. 5(b)), the details of the fine features in the speckle are comparable for $\ell = 0$ and 1, while the structuring is more defined for $\ell = 2$ and 3, so confirming the trend emerging from the FRC curves.

## 4. Conclusions

In this work, we report on a series of single-shot per position ptychography experiments at a seeded FEL source, which were designed to evaluate the effect of the OAM topological charge order $\ell$ on the reconstructed image quality. Our analysis demonstrates an improvement of more than one order of magnitude in spatial resolution and beam stability with respect to previously reported experiments at FEL sources [45,46]. We observe a clear trend relating the achieved image resolution with $\ell$: the higher the $|\ell|$ value, the better the image resolution and quality (also considering the improvement of image contrast at low spatial frequencies). The higher topological charge orders ($|\ell| \geq 2$) show an improvement of about 30% in image resolution with respect to conventional Gaussian beam illumination, paving the way for achieving super-resolution with OAM beams in the EUV regime. Moreover, the beam characterization, performed through the ptychographic reconstruction of the illumination functions, confirms that OAM beams are very sensitive to optical aberrations, which distort the beam transversal profile with respect to ideal Laguerre-Gaussian intensity patterns. For this reason, the experimentally achieved resolution shows a minimum for $|\ell| = 1$, and we were able to ascribe this apparent anomaly to the reduced speckles contrast at high spatial frequencies, which is then improved by increasing $\ell$. Our findings are relevant for the design of fully single-shot FEL-based ptychographic imaging experiments exploiting OAM light, since the use of high $\ell$ values can maximize the beneficial effects of the non-ideality of the illumination function determined by residual beam aberrations.

**Acknowledgments.** We thank Maya Kiskinova for the critical reading of the manuscript. Part of this work has been supported under the European Research Council (ERC) grant agreement No. 866026 (S-BaXIT). G.F.M. acknowledges funding by the European Union's Horizon 2020 research and innovation program, through the grant agreement No. 851154 (ULTRAIMAGE), from Fondazione Cariplo (NanoFast 2020.2544) and from MUR FARE PiXiE (R207A8MNNJ). C.S.B. was supported by Marie Curie Postdoctoral Fellowship DECIPHER, grant [101067016 HORIZON MSCA-2021-PF-01]. P.V. acknowledges support from the Spanish Ministry of Science and Innovation under the Maria de Maeztu Units of Excellence Programme CEX2020-001038-M and the project PID2021-123943NB-I00 (MICINN/FEDER). T.R., M.S. and M.F. acknowledge support from the French "Agence Nationale de la Recherche" under contract ANR-21-CE30-0037 (HELIMAG).

**Disclosures.** The authors declare no conflicts of interest.

**Data availability.** Data underlying the results presented in this paper are not publicly available at this time but may be obtained from the authors upon reasonable request.

**Supplemental document.** See Supplement 1 for supporting content.

# High-resolution ptychographic imaging at a seeded free-electron laser source using OAM beams: supplemental document

## S1. Retrieved beam and object sizes from ptychography reconstructions

The ptychography reconstructions allow obtaining a quantitative measurement of the beam and object sizes at the sample plane, since the reconstructed pixel size $\delta u$ can be precisely evaluated:

$$\delta u = \frac{\lambda}{2} \frac{1}{\sin\left(\mathrm{atan}\left(\frac{l}{2d}\right)\right)} \approx \frac{\lambda d}{l}, \tag{S1}$$

where $\lambda$ is the wavelength (18.9 nm), $d$ is the sample-to-CCD distance (140 mm), and $l$ is the size of the CCD sensor (27.648 mm). Hence, the corresponding reconstructed pixel size is $\delta u \approx 95.7$ nm.

For fully characterizing the probe beam, it is interesting to have an estimate of the illuminating spot size. Figure S1 shows the reconstructed illumination amplitudes for $\ell = \pm 1, \pm 2, \pm 3$. The radius of the white dashed-line circles corresponds to the radius of maximum amplitude $\rho_{\ell,max}$ generated at focus by a spiral zone plate (SZP) and calculated according to [1]

$$\rho_{\ell,max} = \gamma_{|\ell|-1,1} \frac{\lambda f}{\pi D}, \tag{S2}$$

where $\gamma_{|\ell|-1,1}$ is the first root of the $(|\ell|-1)$th-order Bessel function of the first kind, $\lambda$ is the wavelength, $f$ is the focal length (164.7 mm), and $D$ is the diameter of the SZP (1.92 mm). It turns out the lobes of maximum amplitude in the ptychography reconstructions fall close to the calculated ideal radius of maximum amplitude: 1.24 μm for $\ell = \pm 1$, 1.98 μm, for $\ell = \pm 2$, and 2.65 μm for $\ell = \pm 3$.

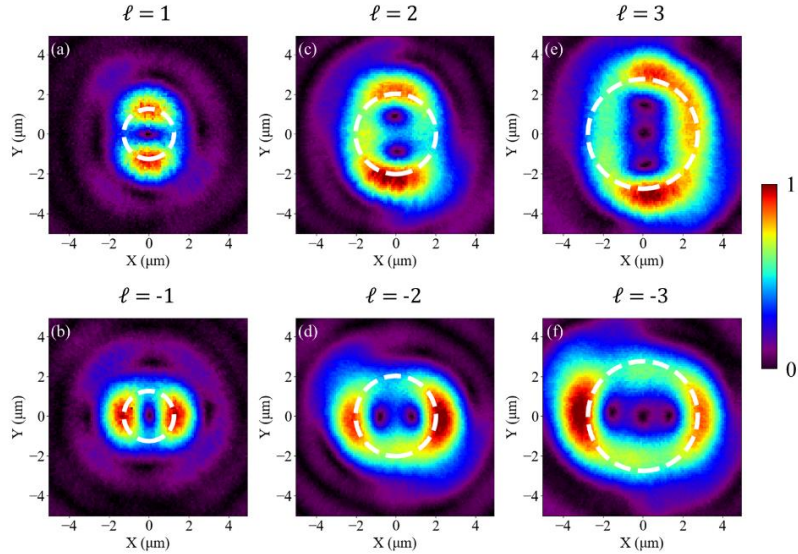

Fig. S1. (a) Reconstructed illumination amplitude for $\ell = 1$ with calibrated pixel size. The radius of the dashed-line circle corresponds to the radius of maximum amplitude obtained from Eq. (S2). (b)-(f) Same reconstructions as in (a) for $\ell = -1, \pm 2, \pm 3$.

**S2. Illumination modes from ptychography reconstructions**

To highlight the relevance of the three $p_i$ illumination modes for each ptychography reconstruction, Fig. S2(a) shows the modes' amplitude for $\ell = 0, 1, 2, 3$, while Fig. S2(b) shows the numerical propagation of those modes to the detector plane via Fraunhofer integral [3]. Evidently, the main modes ($p_1$) contain both the orbital angular momentum illumination and the signature of the optical elements upstream of the SZP (i.e., the central beam stop and the 2 mm input aperture of the DiProI experimental chamber). On the other hand, the intensities of the $p_2$ and $p_3$ modes are confined to the outer part of the reconstructed region, as also observed by Enders et al. [2]. This result can be understood by looking at the propagation of the $p_2$ mode, where the footprint of the used square-shaped order sorting aperture (downstream of the SZP) is visible. By transferring amplitude to the outer mode areas, the reconstruction algorithm can properly handle the presence of diffraction-uncorrelated intensity in the recorded diffraction patterns, without having to correlate it with features in the object transmission function [2].

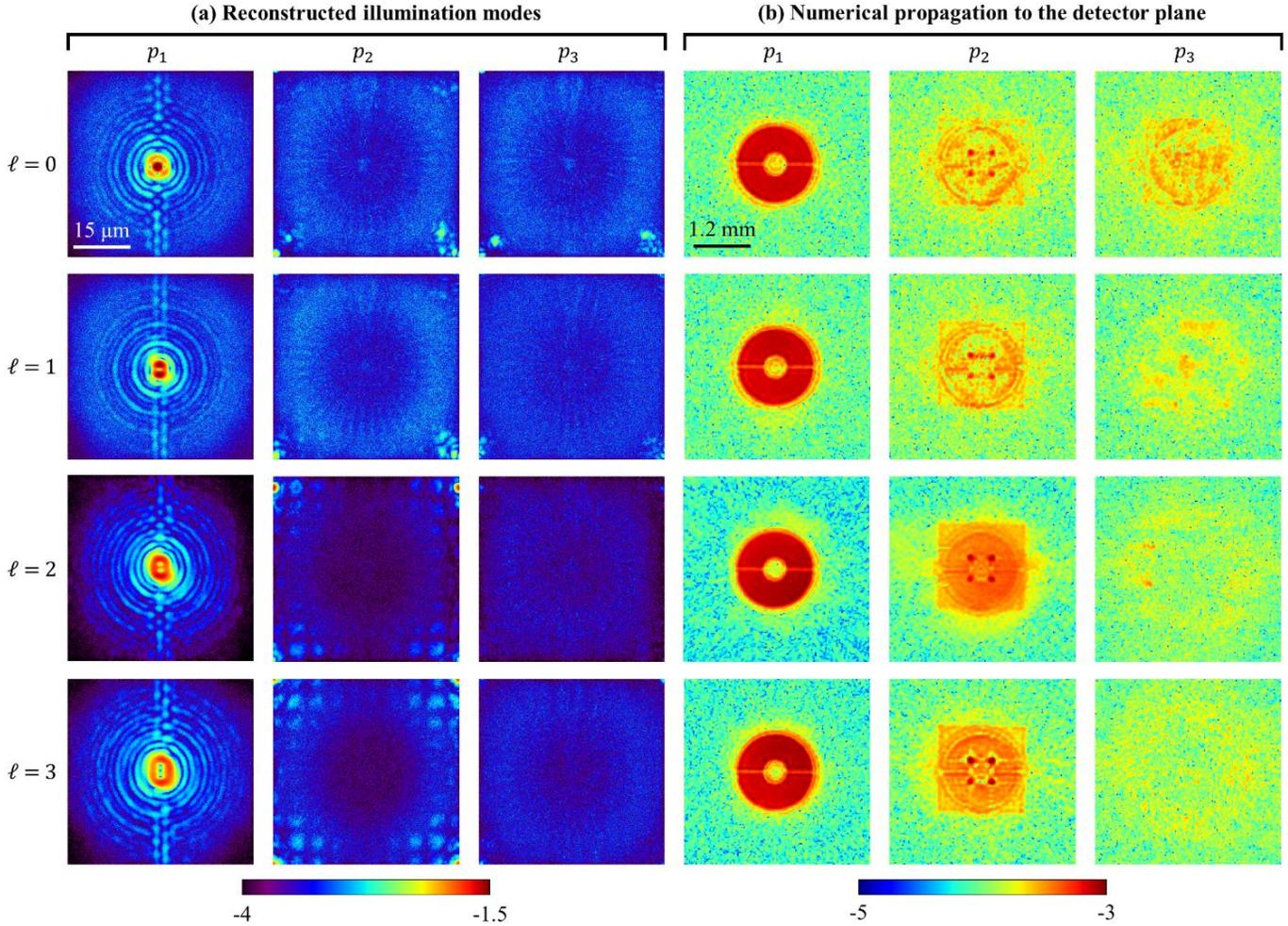

Fig. S2. (a) Modal decomposition of the $\ell = 0, 1, 2, 3$ illumination functions corresponding to some of the ptychography reconstructions in Fig. 2 of the main article. Only the amplitude is shown, in logarithmic scale. (b) Numerical propagation of each $p_i$ mode to the detector plane (140 mm downstream of the sample plane). The patterns are shown in logarithmic scale to highlight the features in the $p_2$ and $p_3$ columns.

## S3. Effect of astigmatism on simulated datasets

To have a qualitative understanding of the impact of astigmatism in our experiment, we simulated the focal spot generated by a $\ell = -1$ SZP with different illuminating wavefronts [4–8]. In order to mimic the experimental conditions, a SZP of focal length $f = 164.7$ mm at $\lambda = 18.9$ nm was considered. To speed up the computation time, the SZP area was limited to a total diameter $D = 0.8$ mm (corresponding to an outermost zone width $dr = 4$ µm). The central part of the illuminating beam was masked by a central beam stop (300 µm in diameter), and the emerging electric field from the SZP was numerically propagated via Fresnel integral [3] to the sample plane.

The SZP was illuminated with five different fields, as summarized in the table of Fig. S3: (a) a spherical wave with curvature radius $r_t = r_s = 17$ m (comparable with the Hartmann wavefront sensor measurements performed during the experiment on the direct beam), (b) an elliptical wavefront with curvature radii $r_t = 17$ m and $r_s = 12.75$ m, (c) an elliptical wavefront with curvature radii $r_t = 17$ m and $r_s = 8.5$ m, and the same elliptical wavefront rotated by (d) $\vartheta = 22.5°$ and by (e) $\vartheta = 45°$ around the propagation direction. To be noticed that the usage of two different curvature radii $r_t, r_s$ accounts for an astigmatic wavefront, and $\vartheta$ is the astigmatism angle with respect to the optical system (i.e. the detector, which defines the x- and y-axis). The curvature radii are not referred to the x- and y-axis when $\vartheta \neq 0°$, hence the "tangential" ($t$) and "sagittal" ($s$) labelling. The amplitude and phase of these configurations at a distance $f$ from the SZP are shown in Fig. S3(a)-(e). In case of a spherical wave, Fig. S3(a), the amplitude displays the ideal "donut-like" shape, and the phase wraps continuously around $2\pi$. In case of an astigmatic wavefront, two bright spots appear in the intensity distribution, and, as it is apparent by comparing Fig. S3(b) and (c), their relative intensity depends on the $r_t/r_s$ ratio of the illuminating wavefront. Figure S3(b),(c) also shows that, for $\vartheta = 0°$, the axis passing through the spots is rotated by $45°$ with respect to the $(x, y)$ reference system. On the other hand, a non-zero astigmatism angle further affects such a rotation angle, as shown in Fig. S3(d),(e). In particular, for $\vartheta = 45°$ (45-degree astigmatism), the two spots lay along the x-axis. This is consistent with the behavior of the reconstructed illumination amplitudes shown in Fig. 2 of the main article.

We conclude with a remark on the role of the topological charge order $\ell$. Figure S4 displays the simulated intensity distributions generated by SZPs as a function of $\ell$ for $r_t = 17$ m, $r_s = 8.5$ m, and $\vartheta = 45°$. Both a variation of the intensity distribution of the focused spots (with the appearance of $|\ell|$ singularities) and an increase in the distance of the peak intensity from the pattern center are visible. In particular, the latter quantity approximately grows as the ideal maximum-amplitude radius $\rho_{\ell,max}$ defined in Eq. (S2) even in the presence of astigmatism, as also confirmed by the experimental measurements in Fig. S1.

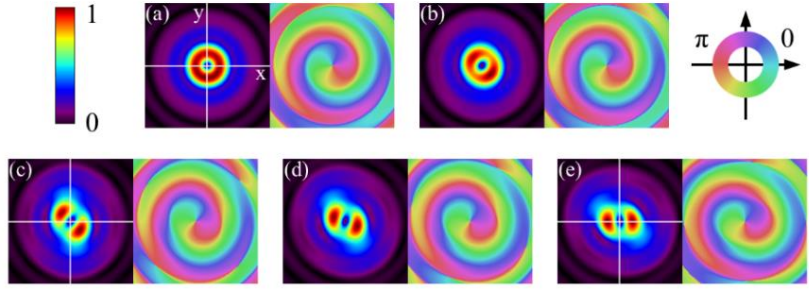

| Panel | $r_t$ (m) | $r_s$ (m) | $\vartheta$ (°) |
|---|---|---|---|
| (a) | 17 | 17 | 0 |
| (b) | 17 | 12.75 | 0 |
| (c) | 17 | 8.5 | 0 |
| (d) | 17 | 8.5 | 22.5 |
| (e) | 17 | 8.5 | 45 |

Fig. S3. (a)-(e) Simulated amplitude (left) and phase (right) of the focal spot generated by a $\ell = -1$ SZP under different illumination conditions. The parameters of the illuminating beams for each case are summarized in the table.

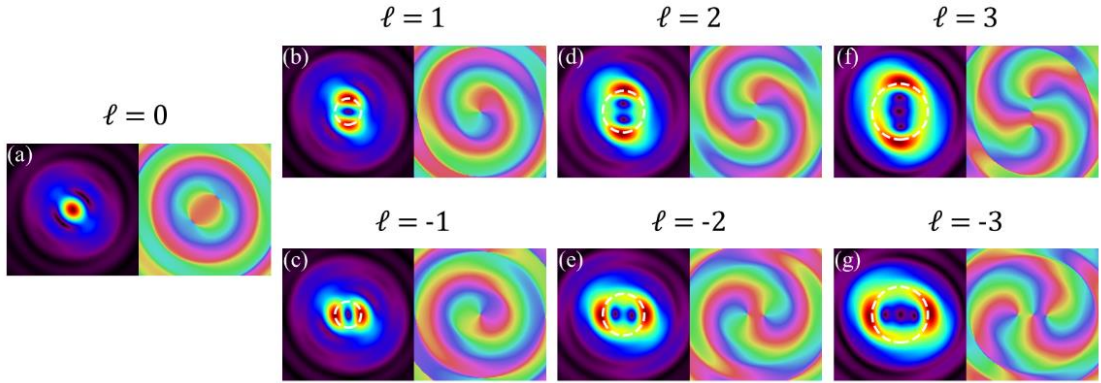

Fig. S4. (a) Simulated amplitude (left) and phase (right) of the focal spot generated by a $\ell = 0$ SZP. The input beam hitting the SZP was characterized by an elliptical wavefront ($r_t = 17$ m and $r_s = 8.5$ m) rotated by $\vartheta = 45°$ around the beam propagation axis. The color bars are the same as in Fig. S3. (b)-(g) Results of similar simulations performed with SZPs for $\ell = \pm 1, \pm 2, \pm 3$. In each case, the reconstructed phase shows $|\ell|$ singularities, as also obtained from the experimental reconstructions in Fig. 2 of the main article. The dashed-line circles represent the maximum-amplitude radius in case of no aberrations, calculated according to the SZPs parameters considered in the simulations.